\documentclass[10pt,conference]{IEEEtran}
\IEEEoverridecommandlockouts
\usepackage{cite}
\usepackage{amsmath,amssymb,amsfonts}
\usepackage{algorithmic}
\usepackage{graphicx}
\usepackage{textcomp}
\usepackage{xcolor}
\def\BibTeX{{\rm B\kern-.05em{\sc i\kern-.025em b}\kern-.08em
    T\kern-.1667em\lower.7ex\hbox{E}\kern-.125emX}}
\graphicspath{{images/}}

\usepackage{balance}

\usepackage{tikz}
\usetikzlibrary{quantikz}
\usepackage{multirow}
\usepackage{pgfplots} 

\usepackage{makecell}

\usepackage{svg}
\graphicspath{{images/}}

\usepackage{orcidlink}
    
\begin{document}

\makeatletter
\newcommand{\linebreakand}{%
  \end{@IEEEauthorhalign}
  \hfill\mbox{}\par
  \mbox{}\hfill\begin{@IEEEauthorhalign}
}
\makeatother

\title{
Can Quantum Computing Improve Uniform Random Sampling of Large Configuration Spaces? 
\thanks{This work has been supported by the German Federal Ministry for Economics and Climate Action in project ProvideQ (FKZ 01MQ22006F) and by the German Ministry for Education and Research in project QuBRA (FKZ:13N16303).}
}
\author{\IEEEauthorblockN{Joshua Ammermann\orcidlink{0000-0001-5533-7274}, Tim Bittner\orcidlink{0000-0001-8608-1689}, Domenik Eichhorn\orcidlink{0000-0001-9428-024X}, Ina Schaefer\orcidlink{0000-0002-7153-761X}}
\IEEEauthorblockA{\textit{Institute of Information Security and Dependability} \\
\textit{Karlsruhe Institute of Technology (KIT)}\\
Karlsruhe, Germany \\
\{name\}.\{surname\}@kit.edu
}
\and
\IEEEauthorblockN{Christoph Seidl\orcidlink{0000-0003-4539-8297}}
\IEEEauthorblockA{\textit{Software Quality Research Group} \\
\textit{IT University of Copenhagen}\\
Copenhagen, Denmark \\
\href{mailto:chse@itu.dk}{chse@itu.dk}
}
}

\maketitle

\begin{abstract}
A software product line models the variability of highly configurable systems.
Complete exploration of all valid configurations (the configuration space) is infeasible as it grows exponentially with the number of features in the worst case. 
In practice, few representative configurations are sampled instead, which may be used for software testing or hardware verification.
Pseudo-randomness of modern computers introduces statistical bias into these samples.
Quantum computing enables truly random, uniform configuration sampling based on inherently random quantum physical effects.
We propose a method to encode the entire configuration space in a superposition and then measure one random sample.
We show the method's uniformity over multiple samples and investigate its scale for different feature models.
We discuss the possibilities and limitations of quantum computing for uniform random sampling regarding current and future quantum hardware. 
\end{abstract}


\section{Introduction}
\label{sec:introduction}
A software product line models the variability of a highly configurable system~\cite{clements2002software}.
It is common for one product line to have thousands of configurable features, which leads to a configuration space  consisting of potentially multiple millions of unique product configurations~\cite{sundermann2020evaluating}.
When new optional features are added to a product line, the size of the configuration space grows exponentially. 
As a result, quality assurance is challenging because exploring all valid configurations is not feasible. 
To overcome the challenge of testing large configuration spaces, practical approaches use algorithms to create samples with few representative configurations. 

\textit{Uniform random sampling} is a technique for sampling random solutions from propositional formulas~\cite{2019-Plazar-UniformSamplingAreWeThereYet, 2017-Oh-FindingNearOptimalConfigs, 2019-Oh-SMARCH}.
Uniform random sampling can be applied to a product line by translating the configuration space to a propositional formula in conjunctive normal form (CNF). 
A CNF consists of variables and clauses, where variables represent the configurable features of a product line, and clauses represent rules stating how features can be combined to create valid product configurations. 
Thus, every solution of the CNF represents one valid product configuration.
A uniform random sample is created by repeatably receiving a random solution of the CNF and thus creating a set of valid product configurations. 
To be uniform, a sampling method must have the property that it selects all solutions with the same probability. 
However, classical uniform random sampling faces two main challenges: 
1$)$ classical computers can only implement pseudo-randomness so that every created sample will have at least a small statistical bias~\cite{vadhan2012pseudorandomness} and 
2$)$ classical algorithms encounter scalability issues as they have to explore exponentially growing configuration spaces to create truly uniform samples~\cite{2019-Plazar-UniformSamplingAreWeThereYet}. 

Quantum computers can solve such issues because they are based on quantum mechanics, for which it is assumed that truly random processes can be executed through superposition and measurement~\cite{deshpande2020implications}.
Furthermore, quantum computers are theoretically proven to provide up to super-polynomial speedups for specific algorithmic problems~\cite{nielsen2002quantum}.

In this paper, we propose a method for quantum-based uniform random sampling of configurations from a software product line using Grover's algorithm~\cite{1996-Grover-Search}. 
Our approach to uniform random sampling on a quantum computer ensures that all solutions are measured with the same probability, making our samples uniform by design.
Additionally, the characteristics of quantum mechanics guarantee that no statistical bias is introduced through pseudo-randomness.
We evaluate this quantum-enabled method for uniform random sampling by showing its uniformity over multiple samples and investigating how it scales for different product lines and configuration spaces. 
We further discuss best and worst-case scenarios for the scalability of the approach. 
In summary, this paper makes the following contributions:
\begin{enumerate}
    \item We provide a method to encode a valid configuration space on a quantum computer to retrieve a random sample.
    \item We provide a detailed explanation and implementation of the quantum algorithm realizing our method for uniform random configuration sampling.
    \item We analyze the uniformity of the generated samples and discuss how current quantum hardware impacts results. 
    \item We evaluate the scalability of our method in terms of the resulting quantum circuit's size. 
    for various product lines.
\end{enumerate}

The content of this paper is structured as follows:
Section~\ref{sec:background} provides background information regarding software product lines, quantum computing, and Grover's algorithm.
Section~\ref{sec:sota} introduces state-of-the-art techniques for uniform random sampling on classical and quantum computers.
Section~\ref{sec:contribution} describes the contribution of this paper: a new method for uniform random sampling that is based on quantum computing using Grover's algorithm.
Section~\ref{sec:eval} evaluates the method for uniform random sampling on multiple example product lines and discusses our insights and results. 
Section~\ref{sec:conclusion} concludes this paper and states possible future work. 

\section{Background}
\label{sec:background}
\subsection{Software Product Lines}

A software product line represents a collection of software artifacts that share a common basis~\cite{clements2002software}. 
Their purpose is to describe the variability of configurable systems from which multiple differing products can be derived. 
One way to describe the configuration space of a product line is by using a \textit{feature model}~\cite{kang1990feature}. 
A feature model represents configuration options as \textit{features}, and it uses propositional logic to define constraints that describe how features can be combined to create valid configurations. 
We show a running example of a feature model representing a simplified version of a car in Fig.~\ref{fig:runningExampleFM}. 
The root feature \textit{car} represents the system's base and includes basic, non-configurable elements. 
All features below \textit{car} define variations of the base systems. 
The features \textit{body, engine} and \textit{gear} are mandatory features, meaning they must be selected when their parent is.
The features \textit{keyless\_entry} and \textit{power\_locks} are optional features.
The features \textit{engine} and \textit{gear} define additional children in specific group types. 
\textit{Electric} and \textit{gas} are combined to an \textit{or group}, and \textit{manual} and \textit{automatic} are combined to an \textit{alternative group}. 
An or group requires that at least one of the included features is selected, whereas an alternative group requires that exactly one of the children is selected. 
Thus, in our example, a car must have one engine type, but it is also valid to select \textit{electric} and \textit{gas} to create a hybrid car. 
However, it is not valid to configure a car with a \textit{manual} and \textit{automatic} transmission. 

The configuration space of the car model constitutes all feature combinations that fulfill the composition rules. 
Thus, a valid configuration is \textit{\{body, engine, gas, gear, manual\}}.
It is invalid because it does not include the mandatory \textit{body} feature.
During the life cycle of a software product line, its feature model is often analyzed to gain empirical or directive information, e.g. to search for unselectable features, create product samples for testing, or count the number of all valid configurations. 
Feature model analysis is automated by transforming the feature model into a propositional formula in conjunctive normal form (CNF). The CNFs are then analyzed with SAT-solver-based methods. For instance, unselectable features can be found by searching for features not contained in any valid configuration. 
Product samples can be created by randomly selecting combinations of features that satisfy the CNF, representing a valid product configuration.  
A minimal CNF that represents the configuration space of the \textit{car} example is the following:
\begin{align*}
(car) \land
(body) \land 
(engine) \land 
(gear)\\
\land~(\neg keyless\_entry \lor power\_locks) \land 
(electric \lor gas)\\
\land(manual \lor automatic) \land 
~(\neg manual \lor \neg automatic)
\end{align*}

\begin{figure}[tb]
    \centering
    \includegraphics[width = \columnwidth]{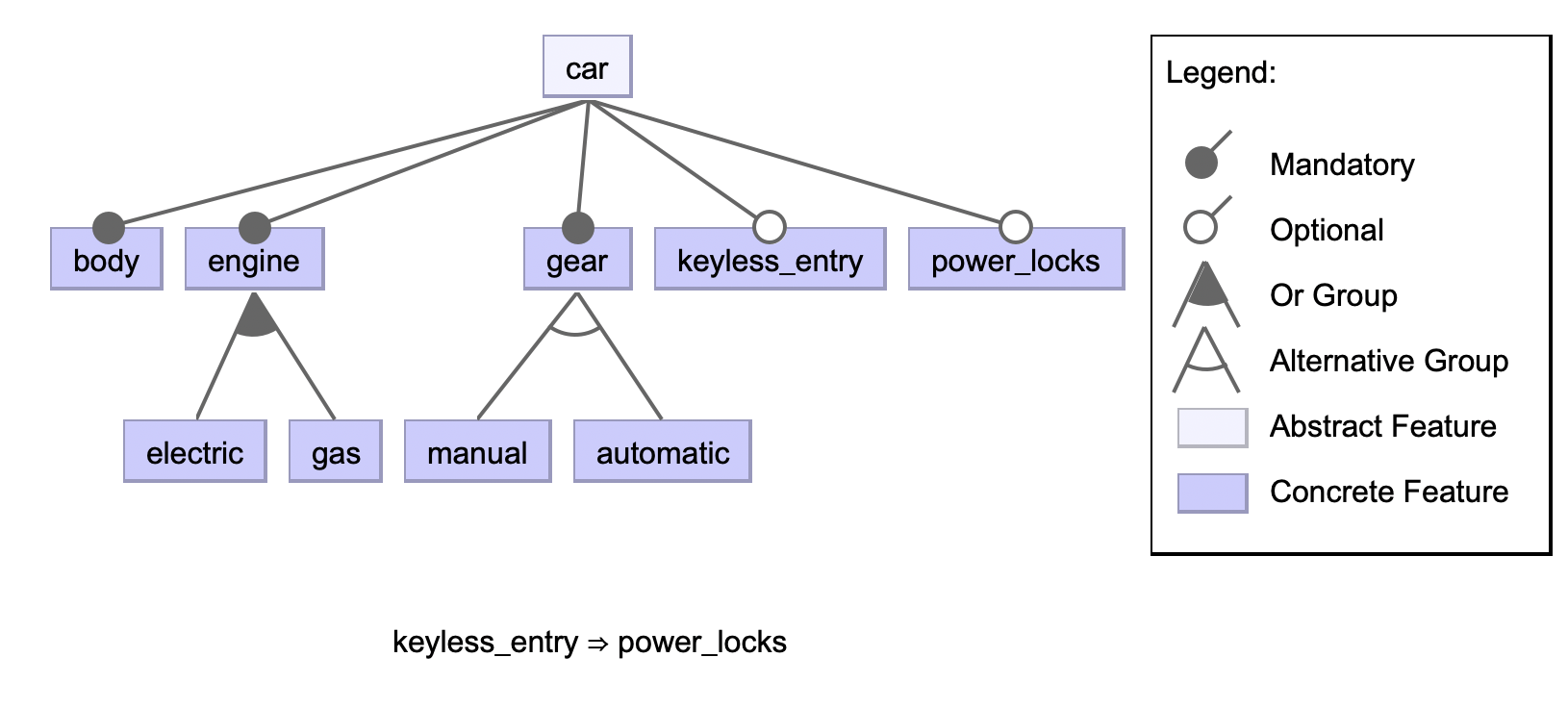}
    \caption{Example feature model representing a car's configuration options.}
    \label{fig:runningExampleFM}
\end{figure}

\subsection{Quantum Computing}
A quantum computer performs calculations on \textit{quantum bits (qubits)} that abstract the state of a quantum mechanical system.
Computations on a quantum computer can leverage quantum mechanical effects, such as superposition and entanglement,  
 to achieve computational advantages over classical computers.

The state of a qubit is described by a complex unit vector $\begin{bmatrix}  a & b \end{bmatrix}^\intercal \in \mathbb{C}^2$.
The \textit{standard/computational basis} of a quantum system consists of the two basis states  
$\ket{0} = 
\begin{bmatrix}
1 & 0 
\end{bmatrix}^\intercal$
and
$\ket{1} = 
\begin{bmatrix}
0 & 1 
\end{bmatrix}^\intercal$.
Any state of a qubit $\ket{\psi}$ can be written as a linear superposition of its two orthonormal basis states $\ket{\psi} = a\ket{0} + b\ket{1}$
where the factors $a$ and $b$ are the probability amplitudes and $|a|^{2}+|b|^{2}=1$ holds.
A geometric representation of a qubit is as a  vector on a Bloch Sphere (see Fig.~\ref{fig:qubitState} (l.)).

\usetikzlibrary{angles, quotes}
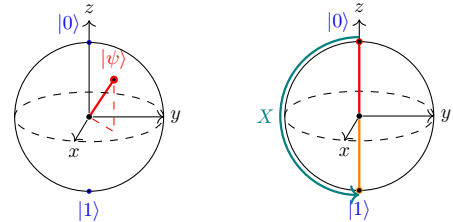
\begin{figure}[b]
    \centering
    \begin{tikzpicture}[scale=0.33, every node/.style={scale=0.75}]]
        \def\r{3}
        
        \draw[draw=red, line width=0.3mm] (0,0) node[circle, fill, inner sep=1] (orig) {} -- (\r/3,\r/2)
        node[circle, fill, inner sep=1, label={[text=red]above:$\ket{\psi}$}, draw=red] (a) {};
        \draw[dashed, draw=red] (orig) -- (\r/3, -\r/5) node (phi) {} -- (a);
        
        \draw (orig) circle (\r);
        \draw[dashed] (orig) ellipse (\r{} and \r/3);
        
        \draw[->] (orig) -- ++(-\r/5, -\r/3) node[below] (x1) {$x$};
        \draw[->] (orig) -- ++(\r, 0) node[right] (x2) {$y$};
        \draw[->] (orig) -- ++(0, 1.3*\r) node[above] (x3) {$z$};

        \draw (0,-\r) node[circle, fill, inner sep=0.7, label={[text=blue]below:$\ket{1}$}, draw=blue] (b0) {};
        \draw (0, \r) node[circle, fill, inner sep=0.7, label={[text=blue]above left:$\ket{0}$}, draw=blue] (b0) {};
            
    \end{tikzpicture}
    \qquad
    \begin{tikzpicture}[scale=0.33, every node/.style={scale=0.75}]]
        \def\r{3}
        
        \draw (orig) circle (\r);
        \draw[dashed] (orig) ellipse (\r{} and \r/3);
        \draw[draw=red] (0,0) node[circle, fill, inner sep=1] (orig) {};
        
        \draw[->] (orig) -- ++(-\r/5, -\r/3) node[below] (x1) {$x$};
        \draw[->] (orig) -- ++(\r, 0) node[right] (x2) {$y$};
        \draw[->] (orig) -- ++(0, 1.3*\r) node[above] (x3) {$z$};
    
        \draw[draw=orange, line width=0.3mm] (orig) {} -- ++(0,-\r);
        \draw (0,-\r) node[circle, fill, inner sep=1, label={[text=blue]below:$\ket{1}$}, draw=orange] (b0) {};
        \draw[draw=red, line width=0.3mm] (orig) {} -- ++(0,\r);
        \draw (0, \r) node[circle, fill, inner sep=1, label={[text=blue]above left:$\ket{0}$}, draw=red] (b1) {};
        
        \pic [draw=teal, angle radius=1.4cm, line width=0.3mm, ->, text=teal, "$X$"'left=0.4] {angle = b1--orig--b0};
    
    \end{tikzpicture}
    \caption{Arbitrary qubit state (l.) and effect of $X$-gate on qubit initially in state $\ket{0}$ (r.) on Bloch Sphere.}
    \label{fig:qubitState}
\end{figure}

A measurement collapses a qubit's state.
When measuring a qubit in state $\ket{\psi}$ (in standard basis):
\begin{itemize}
    \item with probability $|a|^2$ the outcome reads "$0$" and the state changes to $\ket{0}$.
    \item with probability $|b|^2$ the outcome reads "$1$" and the state changes to $\ket{1}$.
\end{itemize}
Relative phases between $a$ and $b$ are not measured but are responsible for interference effects.

To manipulate individual or multiple qubits, arbitrary unitary matrices/transformations $U$ can be applied.
Unitary transformations are reversible so that applying $U$ and its conjugate transpose $U^\dagger$ results in the identity matrix $I$: $UU^\dagger =I$, where $I$ denotes a square matrix with ones on the main diagonal and zeros elsewhere.
The quantum bit-flip operator/gate $X$ (also denoted as NOT and Pauli $X$) rotates a qubit's state by 180° around the $x$-axis of the Bloch Sphere (see Fig.~\ref{fig:qubitState} (r.)).
The $X$ gate behaves like a logical NOT for the computational basis states.
For a qubit in state $\ket{0}$, the $X$ gate flips its state to $\ket{1}$ and vice versa.
The single qubit Hadamard gate $H$ moves a qubit initialized at $\ket{0}$ into the \textit{equal superposition} state $\ket{+}=\frac{1}{\sqrt{2}}\ket{0} + \frac{1}{\sqrt{2}}\ket{1}$, which measures to "$0$" and "$1$" with equal probability of $\frac{1}{2}$. 
The polar basis of a quantum system consists of the two orthonormal basis states $\ket{+}$ and $\ket{-}=\frac{1}{\sqrt{2}}\ket{0} - \frac{1}{\sqrt{2}}\ket{1}$.

A multi-qubit system of $n$ qubits is a quantum system of dimension $d = 2^n$, e.g., with $n=2$ the system has four computational basis states $\ket{00}, \ket{01}, \ket{10}, \ket{11}$, and its state can be described as $\ket{\psi} = \psi_{00}\ket{00} + \psi_{01}\ket{01} + \psi_{10}\ket{10}+ \psi_{11}\ket{11}$.
If one has a two-qubit system with qubits in states $\ket{\alpha}$ and $\ket{\beta}$ the overall system's state can be described by a so-called product state $\ket{\psi} = \ket{\alpha}\otimes\ket{\beta}$. 
Qubits can be \textit{entangled} by which they correlate in a way that, when measured, the entangled qubits collapse to the same state $\ket{0}$ or $\ket{1}$.
Due to the entanglement, such an entangled state can not be written as a product state.
The controlled-$X$ gate $CX$ is a two-qubit gate that flips the second qubit (target) if and only if the first qubit (control) is in $\ket{1}$.
It is used to create entanglement.

\textit{Quantum circuits} are used to model the behavior of a quantum system over time.
Typically, a quantum circuit defines qubits, their initial states, the application of specific quantum gates, and the measurement of qubits.

In practice, quantum computers only can realize a small set of gates.
The gate set available in many quantum computers is universal, s.t. every other transformation $U$ can be constructed from the physically available gates.
A quantum computer also has a topology that defines the physical connections between qubits and thus restricts their possible interactions (e.g., only specific qubits can be entangled on a specific quantum computer).
A quantum circuit, thus, has to be transpiled to match a quantum computer's available gates as well as the computer's
topology.

The $X$ gate behaves like a logical NOT as described in the background.
Classical AND and OR are not reversible, but they can be simulated on a quantum computer by using an additional qubit that is initialized to $\ket{0}$ by default~\cite{2011-Williams-book}.
Such auxiliary qubits are called ancilla qubits and are required to perform these operations.
Williams~\cite{2011-Williams-book} showed how classical AND and OR with two inputs can be simulated on a quantum computer, which we extend to work on $n$ inputs.
Tab.~\ref{tab:transformation} shows how logical AND and OR can be realized on a quantum computer in the form of
an exemplary circuit fragment.
An AND operation as defined by $U_{and}$ applies a bit-flip ($X$ gate) on the target ancilla if all $n$ input qubits are in state $\ket{1}$. 
Thus, it can be realized using a multi-controlled-$X$ gate.
A multi-controlled-$X$ gate for $n=2$ is a Toffoli-Gate.
A quantum OR operation can be derived from AND and NOT operations similar to classical Boolean algebra.
First, every single qubit is flipped using $X$-gates, then the $U_{and}$ operation is applied, then all qubits except the target are flipped again using $X$-gates:
$U_{or} = X^{\otimes n} \, U_{and} \, (X^{\otimes n-1} \otimes I)$.

\newsavebox{\boxA}
\savebox{\boxA}{
\begin{tikzpicture}
        \node[scale=0.8] {
            \begin{quantikz}
            a & \gate{X} & \qw \neg a
        \end{quantikz}
    };
\end{tikzpicture}}

\newsavebox{\boxB}
\savebox{\boxB}{
\begin{tikzpicture}
    \node[scale=0.7] {
        \begin{quantikz}
        a\;  & \ctrl{1} & \qw \; a \\
        \vdots\; & \ctrl{1} \vqw{1} & \qw \;\vdots \\
        n\;  & \ctrl{1} \vqw{1} & \qw \; n \\
        \ket{0}\; & \targ{} & \; \qw a \land \dots \land n
        \end{quantikz}
    };
\end{tikzpicture}}

\newsavebox{\boxC}
\savebox{\boxC}{
\begin{tikzpicture}
    \node[scale=0.7] {
        \begin{quantikz}[row sep=0.2cm]
        a\; & \gate{X} & \gate[wires=4]{U_{and}} & \gate{X} & \qw \; a \\
        \vdots\; & \gate{X} & & \gate{X} & \qw \;\vdots \\
        n\; & \gate{X} & & \gate{X} & \qw \; n \\
        \ket{0} & \gate{X} &  & \qw & \; \qw a \lor \dots \lor n
        \end{quantikz}
    };
\end{tikzpicture}}

\setlength{\extrarowheight}{3pt}
\begin{table}[bt]
    \centering
    \caption{Logical AND and OR as quantum circuit fragments}
    \begin{tabular}{|c|c|c|}
        \hline
        \textbf{AND} & \textbf{OR} \\
        \hline
        \makecell[c]{\usebox\boxB} &
        \makecell[c]{\usebox\boxC}
        \\\hline
    \end{tabular}
    \label{tab:transformation}
\end{table}

\subsection{Grover's Algorithm}
\label{sec:grover}
We're using Grover's algorithm~\cite{1996-Grover-Search} for quantum search over unstructured solution sets to sample a configuration space uniformly random (see Section~\ref{sec:contribution}).
Let $f(x)$ be a black box function with an integer $x$ as input which can validate a solution ($f(x)=1$) for a specific search problem.
Given such a function $f(x)$ the algorithm finds a unique solution $m$ from the set of all solutions $M$~\cite{2016-Nielsen-QCandQI}.
The size of the overall search space $N$ is $2^n$ and the index of elements in the search space can be stored in $n$ bits classically~\cite{2016-Nielsen-QCandQI}.
With high probability, a solution is found using at most $O({\sqrt{N/M}})$ calls to $f(x)$ instead of $O(N/M)$ calls necessary classically~\cite{2016-Nielsen-QCandQI}.

Detailed algorithm analysis is conducted in~\cite{2016-Nielsen-QCandQI}, which we summarize in the following.
Onto $n$ qubits that are initial in $\ket{0}$ state, the Hadamard gate $H$ is applied to every qubit.
This results in the equal superposition state ${\ket{s}=\frac{1}{\sqrt{N}}\sum_{x=0}^{N-1}\ket{x}}$ over the search space $N$.
Then, a Grover iteration is applied $k=\mathcal{O}(\sqrt{N/M})$ times, where the number of iterations is denoted by $k$.

An iteration consists of the oracle operator $U_\omega$ first and the diffusion operator $U_s$ afterward.
$U_\omega$ marks the solutions by shifting their phases, as it can - by the nature of being an oracle - recognize the solutions of the problem.
The oracle is treated as a black box to generalize the algorithm, as the oracle encapsulates a problem-specific structure.
In contrast, the diffusion operator is defined generally as ${U_s = H^{\otimes n}(2\ket{0}\bra{0}-I)H^{\otimes n} = 2\ket{s}\bra{s}-I}$ and performs a reflection over $\ket{s}$ to amplify the amplitudes of the marked solutions.

\begin{figure}[tb]
    \centering
    \begin{tikzpicture}[scale=0.7, every node/.style={scale=0.75}, baseline={(0,-0.6)}]]
        \def\r{3}
        
        \draw (0,0) node[circle, fill, inner sep=1] (orig) {};
        
        \draw[->] (orig) -- ++(0, \r) node[above] (x1) {$\ket{w}$};
        \draw[->] (orig) -- ++(\r, 0) node[right] (x2) {$\ket{s'}$};

        \draw[->] (orig) -- ++(0.92387953*\r, 0.38268343*\r) node[label={right:$\ket{s}$}] (s) {};
        \draw[->] (orig) -- ++(0.92387953*\r, -0.38268343*\r) node[label={right:$U_\omega \ket{s}$}] (s2) {};
        \draw[->] (orig) -- ++(0.38268343*\r, 0.92387953*\r) node[label={right:$U_s U_\omega \ket{s}$}] (s3) {};

        \pic [draw=teal, angle radius=1.8cm, <-, text=teal, "$U_\omega $"'{right=0.8, anchor=north}] {angle = s2--orig--s};
        \pic [draw=cyan, angle radius=2.5cm, ->, text=cyan, "$U_s $"{right=0.8, anchor=west}] {angle = s2--orig--s3};
        
        \pic [draw=blue, angle radius=1.5cm, text=blue, "$\theta/2$"{scale=0.8, right}] {angle = x2--orig--s};
        \pic [draw=blue, angle radius=1.5cm, text=blue, "$\theta/2$"{scale=0.8, right}] {angle = s2--orig--x2};
        \pic [draw=blue, angle radius=1.4cm, text=blue, "$\theta$"{right}] {angle = s--orig--s3};
            
    \end{tikzpicture}
    \caption{Geometric interpretation of the first iteration of Grover's algorithm.}
    \label{fig:grover}
\end{figure}
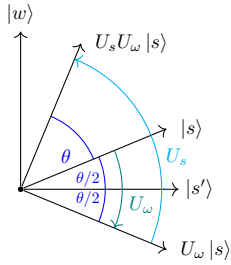

Fig.~\ref{fig:grover} shows the geometric interpretation of the first iteration of Grover's algorithm in the two-dimensional subspace spanned by \ket{\omega} and $\ket{s'}$.
$\ket{\omega}$ denotes the equal superposition over all $x$ that are solutions to the search problem (the 'winning' states), and $\ket{s'}$ denotes the equal superposition over all $x$ that are not the solution to the search problem (the 'losing' states).
The initial equal superposition state can be rewritten as $\ket{s} = \sqrt{\frac{N-M}{N}}\ket{s'}+\sqrt{\frac{M}{N}}\ket{\omega}$ in this subspace.
Geometrically, the oracle operator $U_\omega$ is a reflection of $\ket{s}$ over $\ket{s'}$.
Then the diffusion operator $U_s$ reflects $U_\omega\ket{s}$ over $\ket{s}$, resulting in an overall rotation of $\theta$ towards $\ket{\omega}$ per iteration.
If $M \leq \frac{N}{2}$ then $\sin{\frac{\theta}{2}}={\sqrt{\frac{M}{N}}}$ holds.
After $k$ iterations of the oracle and diffusion operators, a rotation of $2k\theta$ was applied to $\ket{s}$, and the amplitudes of the winning states are continuously amplified.
Finally, the $n$ qubits are measured to obtain a searched element with high probability.

The number of iterations that first yields a near-optimal solution (closest to $\ket{\omega}$) is denoted as $k_{best}\leq \left\lceil\frac{\pi}{4}\sqrt{\frac{N}{M}}\right\rceil$.
The quantum counting algorithm~\cite{1998-Brassard-QuantumCounting} is based on quantum phase estimation and Grover's algorithm and estimates the value of $\theta$, from which $k_{best}$ can be calculated.
Grover's algorithm with $k_{best}$ iterations results in a state that is closest to $\ket{\omega}$, 
so the probability amplitudes of all solution states are $\approx \frac{1}{\sqrt{M}}$ and the amplitudes of the other states are $\approx 0$.
Iterations of Grover's algorithm can not be applied partially, 
so $k_{best}$ has to be rounded up to the next integer (thus the ceiling operator).
This means that after applying Grover's algorithm with $k_{best}$ iterations,
still a small probability of measuring a wrong state remains (except for a small number of problem instances where $k_{best}$ is not rounded up).

A total of $log(N)$ gates are required to implement the diffusion operator $U_s$ so the overall complexity is $O(\sqrt{\frac{N}{M}}log(N))$ without the cost for the construction of the oracle $U_\omega$~\cite{2016-Nielsen-QCandQI}.
Grover's speedup is quadratic and not exponential but still considerable for large $N$.
Grover's algorithm was proven to be optimal in oracle calls in~\cite{1999-Zalka-GroverIsOptimal}.

\section{State of the Art}
\label{sec:sota}
We provide an overview of state-of-the-art classical and quantum sampling techniques.

\subsection{Classical Uniform Random Sampling}
Uniform random sampling is an active field of research with various sampling approaches. 
Available samplers can be categorized into heuristic-~\cite{2018-Dutra-QuickSampler, 2021-Golia-CMSGen}, hashing-~\cite{2015-Chakraborty-UniformSatWitnessGeneration, 2020-Soos-UniGen3} and counting-based~\cite{2018-Sharma-KUS, 2018-Achlioptas-SPUR, 2019-Oh-SMARCH, 2022-Heradio-UniformAndScalableSAT-Sampling} samplers~\cite{2022-Heradio-UniformAndScalableSAT-Sampling}.

QuickSampler~\cite{2018-Dutra-QuickSampler} uses atomic mutations and heuristics for uniform random sampling. 
Another heuristic sampler, CMSgen~\cite{2021-Golia-CMSGen}, uses an SAT solver based on conflict-driven cause learning.
UniGen2~\cite{2015-Chakraborty-UniformSatWitnessGeneration} is a random hashing-based sampler that supports parallelization and it's runtime was improved in UniGen3~\cite{2020-Soos-UniGen3}.
KUS~\cite{2018-Sharma-KUS}, SPUR~\cite{2018-Achlioptas-SPUR}, SMARCH~\cite{2019-Oh-SMARCH} and BDDSampler~\cite{2022-Heradio-UniformAndScalableSAT-Sampling} are counting-based samplers.
Knuth's algorithm~\cite{2009-Knuth-Book} using Binary Decision Diagrams (BDDs) was applied to SPLs by Oh et al. in~\cite{2017-Oh-FindingNearOptimalConfigs}.
KUS uses the Deterministic Decomposable Negation Normal Form (d-DNNF) - a superset of ordered BDDs~\cite{2018-Sharma-KUS}.
SPUR and SMARCH rely on Thurley's \#-SAT solver~\cite{2006-Thurley-sharpSAT} which counts the number of satisfying assignments of a propositional formula (which are valid configurations in our case)~\cite{2018-Achlioptas-SPUR, 2019-Oh-SMARCH}.
BDDSampler uses ordered and reduced BDDs~\cite{2022-Heradio-UniformAndScalableSAT-Sampling}.

Comparisons of uniform random samplers showed a trade-off between scalability and uniformity~\cite{2018-Achlioptas-SPUR, 2018-Dutra-QuickSampler, 2018-Sharma-KUS, 2019-Chakraborty-Barbarik, 2022-Soos-QuantitativeTestingOfSamplers, 2019-Plazar-UniformSamplingAreWeThereYet, 2020-Heradio-UniformAndScalableSAT-Sampling, 2022-Heradio-UniformAndScalableSAT-Sampling}.
To the best of our knowledge, the time complexity for uniform random samplers is not fully understood yet and is an open question.
The results of Heradio et al.~\cite{2022-Heradio-UniformAndScalableSAT-Sampling} hint that additional aspects to the number of variables of the CNF influence sampling time, such as the number of clauses and complexity, and indicate an overall exponential growth.

\subsection{Quantum Sampling}

To the best of our knowledge, no other work discusses uniform random sampling using quantum algorithms in the way this paper does.
Hangleiter and Eisert~\cite{2022-Hangleiter} elaborate different random sampling schemes for quantum circuits and the computational advantage of using quantum computers for this.
They define sampling from a random quantum computation as quantum random sampling.
We are not interested in random quantum computations, but in (uniform) random samples from a constrained distribution.
These constraints can not be enforced by mere randomness, so our work does not relate to theirs at all.

\section{Uniform Random Sampling of Configurations with Grover's Algorithm}
\label{sec:contribution}
This section shows our method to use Grover's search algorithm to sample a configuration space of a configurable system uniformly.
\begin{figure}[b]
    \centerline{\includegraphics[
    width=0.5\textwidth]{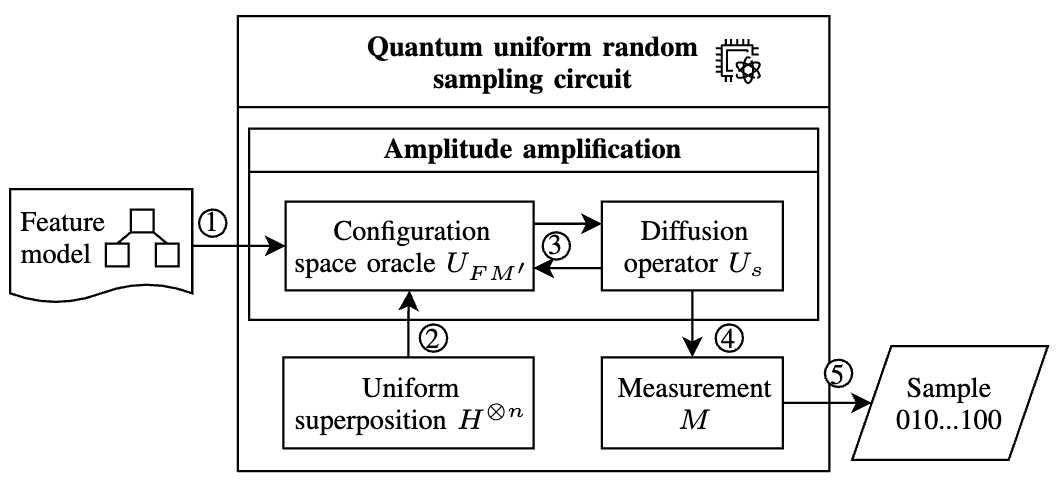}}
    \caption{Overview of our method for uniform random sampling that encodes a specific feature model into a configuration space oracle, which is used for amplitude amplification to obtain a uniform random sample.}
    \label{fig:overview}
\end{figure}
\begin{figure*}[bt]
    \centering
    \begin{tikzpicture}
    \node[scale=0.8] {
        \begin{quantikz}[row sep=0.1cm]
            q_7 (manual)\; & \gate[5, nwires={3,4}]{U_{or}} \gategroup[5,steps=1, background, style={dashed, rounded corners,fill=blue!20, inner xsep=2pt}]{{$a_7 = q_7 \lor q_8$}} 
                & \gate{X} \gategroup[6,steps=3, background, style={dashed, rounded corners,fill=blue!20, inner xsep=2pt}]{{$a_8 = \neg q_7 \lor \neg q_8$}} & \gate[6, nwires={3,4,5}]{U_{or}}  & \gate{X}
                & \qw & \ldots &
                & \gate{X} \gategroup[7,steps=4, background, style={dashed, rounded corners,fill=blue!20, inner xsep=2pt}]{{\sc Uncompute}}
                & \gate[6, nwires={3,4,5}]{U_{or}\dagger} & \gate{X}
                & \gate[5, nwires={3,4}]{U_{or}\dagger} & \qw \\
            q_8 (automatic) \; &  & \gate{X} & & \gate{X} & \qw & \ldots &  & \gate{X} & & \gate{X} & & \qw \\
            \vdots\; & & & & & & & & & & & & \vdots\\
             & & & & & \ldots & \gate[4]{U_{and}} \gategroup[4,steps=1, background, style={dashed, rounded corners,fill=blue!20, inner xsep=2pt}, label style={label position=below,anchor=north,yshift=-0.2cm}]{{$...\land a_7 \land a_8$}} & \qw\ldots & & & & &  \\
            a_7\; & & \qw\ldots & & & \ldots &
                & \ldots\qw & & & \ldots & \qw & \qw \\
            a_8\; & \qw & \qw & \qw & \qw & \qw\ldots & & \ldots\qw & \qw & \qw & \qw  & \qw & \qw \\
            q_{tar}\; & \gate{X} \gategroup[1,steps=2, background, style={dashed, rounded corners,fill=blue!20, inner xsep=2pt}, label style={label position=below,anchor=north,yshift=-0.2cm}]{{\sc Initialize $\ket{-}$}}
                & \gate{H} & \qw & \qw & \qw\ldots & & \ldots\qw & \qw & \qw & \gate{H} & \gate{X} & \qw
        \end{quantikz}
    };
    \end{tikzpicture}
    \caption{Excerpt of the configuration space phase oracle $U_{FM'}$  for the running example with 10 variables and 8 clauses.}
    \label{fig:satCircuit}
\end{figure*}
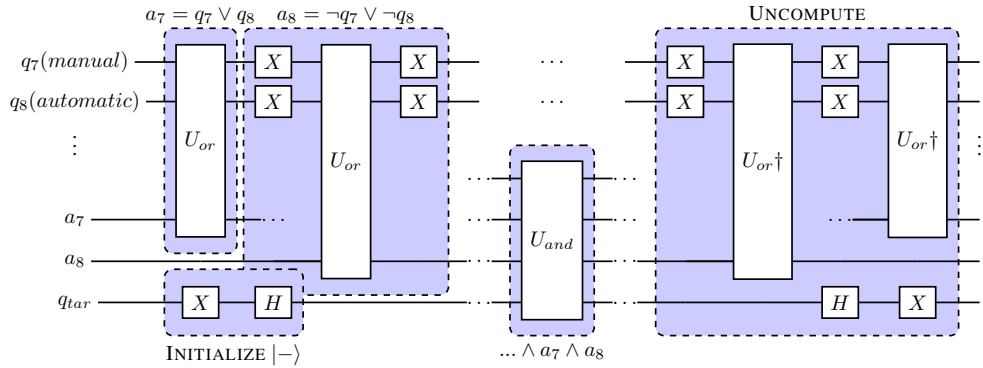
Fig.~\ref{fig:overview} provides an overview of our method.
The method starts with converting a feature model into a CNF~(1).
We create a uniform superposition~(2) of all configuration options on a quantum computer and then construct an oracle~(3) that knows about the validity of a given configuration, which represents the solution of a CNF. 
We combine the oracle with Grover's algorithm to amplify the solutions of the CNF and then perform a measurement~(4). 
The measurement randomly returns one possible solution of the CNF~(5), which is equal to randomly obtaining a valid configuration of the product line. 
After repeating this process multiple times, we derive multiple random configurations, representing a uniform random sample of the configuration space.

\subsection{Creating a Configuration Space Oracle}

We introduced quantum circuit fragments that are analogous to classical AND, OR, and NOT gates in our Background Section~\ref{sec:background} 
Using these three circuit fragments a quantum circuit for an entire CNF formula can be defined.
This circuit requires one qubit $q$ for each input variable in the CNF, one ancilla qubit $a$ for each clause in the CNF, and one output qubit $q_{tar}$.
First, each clause $c$ of the SAT formula is transformed into a circuit fragment by
applying $U_{or}$ controlled by each qubit $q$ present in the clause.
$U_{or}$ targets a separate $a_c$ and $X$-gates are applied before and after $U_{or}$ on $q$ if the variable is negated in $c$.
We denote the transformation of all separate clauses as $U_{clauses}$.
After applying $U_{clauses}$ once, one $U_{and}$ is applied controlled by all ancillas $a$ onto $q_{tar}$.
Finally, to uncompute the qubit states, the conjugate transpose ${U_{clauses}}^\dagger$ is applied.
This results in the feature model oracle $U_{FM}$.

$U_{FM}$ can predict if a variable assignment encoded into the input qubits is valid - specifically a valid configuration - on the target qubit ($q_{tar}=\ket{1}$ if the assignment is valid).
Furthermore, if the input qubits are initialized in uniform superposition over the whole configuration space, only valid configurations can be derived with a high probability using Grover's Algorithm.

\subsection{Applying Grover's Algorithm using the Oracle}
To use $U_{FM}$, which marks $q_{tar}$ with $\ket{0}$ or $\ket{1}$, it has to be transformed into a phase oracle that instead marks the phases of states.
Such a behavior can be achieved by a phase kickback onto $q_{tar}$.
For this, a sequence of $XH$ gates has to be added before $U_{FM}$ on $q_{tar}$.
After $U_{FM}$, the inverse sequence $HX$ has to be added as well.
This transforms $q_{tar}$ into the $\ket{-}$ state before applying $U_{FM}$ and reverses the transformation afterward.

Fig.~\ref{fig:satCircuit} displays an excerpt of the phase oracle $U_{FM'}$ for the running example:
For this problem instance with 10 variables and 8 clauses, the oracle has 19 qubits (10 input variable qubits + 8 ancilla qubits + 1 target qubit).
Fig.~\ref{fig:satCircuit} shows two wires representing variables the $manual(q_7)$ and $automatic(q_8)$ of the example CNF.
These variables are used in two clauses, which are transformed into $U_or$ and $X$ gates and connected to the ancilla qubits $a_7$ and $a_8$ respectively.
The qubit $q_tar$ for the results of the conjunction is initialized in the $\ket{-}$ state as discussed above.
After applying the conjunction ($U_{and}$) on all ancillas, the rest of the circuit is uncomputed so that only the phase information applied by $U_{and}$ remains.

The configuration space phase oracle $U_{FM'}$ is then used in Grover's Algorithm as oracle operator $U_\omega$ (see Section~\ref{sec:grover}).
The valid configurations are marked by $U_{FM'}$ and amplified by the diffusion operator $U_s$.
With this, the circuit construction is completed and can be deployed to simulators or quantum hardware.

\section{Evaluation}
\label{sec:eval}
We investigate the current state of quantum computing for uniform random sampling of large configuration spaces and give predictions about future prospects. 
Due to the limitations of current quantum hardware and simulation of such, executing quantum-enabled uniform random sampling and comparing its empirical results to state-of-the-art classical random sampling algorithms will not lead to reasonable results.
Any instances of uniform random sampling problems that could be run on current quantum hardware are solved easily using classical algorithms, eliminating the need for quantum computations.
Hence, to judge the performance of quantum-enabled random sampling for relevant problem sizes, we perform a theoretical analysis.
As the computational complexity of Grover's search has already been determined to be $O(\sqrt{N})$, our algorithm conforms to this bound.
We consider three research questions about the feasibility, uniformity, and scale of our approach.
We describe the methodology to answer these questions, present the results and interpret them in a discussion.

\subsection{Research Questions}
We pose the following research questions for the evaluation of our approach: 
\subsubsection{Feasibility} 
Is it possible to construct quantum circuits that encode the valid configuration space and apply Grover's search algorithm to retrieve a valid configuration when measuring?
\subsubsection{Uniformity} 
If one measures the output multiple times to retrieve multiple configurations, how uniformly distributed are the obtained samples of the configuration space?
\subsubsection{Scalability} 
How do runtime and hardware requirements scale in relation to the size of the (valid) input configuration space?

\subsection{Methodology}

To answer our questions, we implemented the proposed method (see Section~\ref{sec:contribution}) that transforms a given feature model from a Boolean CNF into a quantum circuit. 
The method was implemented as a Jupyter Notebook using Qiskit.
It and the scripts created for this evaluation are available on GitHub\footnote{\url{https://github.com/KIT-TVA/grover-uniform-random-sampling}}.
This approach allows us to reason about feasibility and uniformity.
Feasibility can be evaluated by the fact that we are able to successfully implement and execute the proposed algorithm.
While, in principle, uniformity could be assessed by analyzing the measured outputs of the generated quantum circuits, either in a simulator or on real hardware, current limitations do not allow a meaningful empirical investigation:
Simulators are not capable of executing larger circuits with many qubits, and current NISQ quantum hardware inherently leads to errors.
While we cannot run or simulate the generated circuits, we are able to construct and reason about them without execution. Hence, we investigate circuits for the feature models of the BURST benchmarking suite~\cite{2021-Archer-BURST} as well as feature models provided by the software product line tool FeatureIDE~\cite{FeatureIDE}.
For the investigation of scalability, we observe quantum circuit metrics such as the number of qubits required (width) and the number of gates required (depth) for a specific configuration space.
This also allows us to estimate how capable current-generation quantum computers are of dealing with these problem instances.

\subsection{Results}

Tab.~\ref{tab:fmanalysis} shows the results of our circuit construction and analysis.
In the left half, we present metrics regarding the feature model, e.g., the number of features and the number of clauses when transformed to CNF.
The transformation from feature model to CNF is not part of our work but we noticed that the available DIMACS files are not optimized, e.g., some clauses are repeated, artificially bloating the size of the CNF.
We opted to not optimize the CNF formulas ourselves but take the input as is to be consistent with the real-world data.
\begin{table*}[tb]
    \caption{Results of the Evaluation for Different Feature Models}
    \begin{center}
    \begin{tabular}{|l|c|c|c|c||c|c|c|}
        \hline
        \multirow{3}{8em}{\textbf{Feature model}} & \multirow{3}{4em}{\textbf{\# of features}} & \multirow{3}{4em}{\textbf{\# of clauses}} & \multirow{3}{5em}{\textbf{\# of valid configurations}} & \multirow{3}{5em}{\textbf{\% valid configurations}} & \multirow{3}{4em}{\textbf{$k_{best}$}} &\multirow{3}{4em}{\textbf{Circuit width}}  & \multirow{3}{11em}{\textbf{Circuit depth transpiled for Statevector backend (IBM) for $k=1$}}\\
        & & & & & & & \\
        & & & & & & & \\ 
        \hline 
        car & 10 & 17 & 1,80e+01 & 1,76\% & 5 & 28 & 162 \\
        sandwich & 19 & 27 & 2,81e+03 & 0,54\% & 10 & 47 & 322 \\
        bike & 54 & 127 & 1,12e+07 & 6,22e-10 & 31502 & 182 & 88 \\
        kconfig\_axTLS & 96 & 183 & 8,26e+11 & 1,04e-17 & 2,43e+08 & 280 & 218 \\
        kconfig\_uClibc & 313 & 1240 & 1,66e+40 & 9,95e-55 & 7,87418e+26 & 1554 & n/a \\
        kconfig\_Busybox\_1.18.0 & 854 & 1163 & 2,06e+201 & 1,72e-56 & 5,99576e+27 & 2018 & n/a \\
        kconfig\_EMBToolkit & 1179 & 5414 & 5,13e+96 & 6,24e-259 & 9,9316e+128 & 6594 & n/a \\
        \hline
    \end{tabular}
    \end{center}
    \label{tab:fmanalysis}
\end{table*}
The number of features and clauses directly contributes to the generated circuit width (column on the right).
Our approach requires one qubit per feature and one ancilla qubit per clause.
Additionally, one extra ancilla qubit is required for the conjunction of all clauses.
This means our circuit width is  $1+\#\mathit{features}+\#\mathit{clauses}$.

We show the circuit depth for one iteration ($k = 1$) for circuits transpiled for the IBM Statevector simulator as we are interested in
the total depth of a quantum circuit to reason about the possibility of running small examples on real quantum hardware. 
However, the circuit depth would increase further when transpiling for actual quantum hardware as not all qubits are connected and SWAP-gate sequences would have to be introduced to allow for the execution of a circuit.
To determine the total simulator circuit depth, it is required to know how many Grover iterations $k_{best}$ have to be performed.
As we have shown in Section~\ref{sec:grover} that $k_{best}$ depends on the number of valid entries in the search space, we determine the number of valid configurations for each feature model using the model counter GANAK~\cite{GANAK}.
Using this number and the fact that our search space is $2^{\#Features}$, we have all data needed to determine $k_{best}$.
The total simulator circuit depth can now be calculated by $k_{best} * (depth_{k=1}  - 1) + 1$.
Subtracting one from the circuit depth before multiplication is necessary as the given depth includes the superposition imposed by a one-deep set of Hadamard gates for initialization, which has to be executed only once.
For the car feature model, the total gate depth amounts to $806$. 
The other total depth values are computed analogously and are in the order of $k_{best}$. 

For each circuit, we also collect the ratio of the valid configuration space compared to the theoretical maximum size of the configuration space, which is exponential in the number of features, i.e., the percentage of valid configurations. 
This data point is collected because it allows us to reason about the runtime of Grover's algorithm for a specific problem class.

\subsection{Discussion}

Based on the results we collected, we now present answers to our research questions.

\subsubsection{Feasibility}
Our results show that it is possible to construct a quantum circuit based on Grover's search to randomly retrieve valid configurations from a configuration space. 
However, when executing or simulating circuits of even the smallest examples, we find that there will likely be incorrect or missing results due to the limitations of current NISQ-era quantum hardware.
However, in the future, when quantum hardware with a large number of error-free qubits and gates is available, the presented approach will be feasible for uniform random sampling. 

\subsubsection{Uniformity}
Assessing uniformity proves to be difficult for the quantum algorithm.
In theory, simulation with IBM Statevector perfectly (uniformly) amplifies the amplitudes of valid configurations while decreasing invalid states' amplitudes.
This would also be the case for a perfectly error-corrected quantum computer.
Yet, there remains the chance to measure an invalid configuration as the number of Grover iterations $k$ has to be rounded.
Rounding $k$ is required as we can not add fractions of our oracle-diffuser iterations to our circuit.
However, the equal superposition start vector $\ket{s}$ rarely requires an integer amount of reflections to get to $\ket{\omega}$ that encodes the valid solutions.\footnote{See the graphical explanation of Section~\ref{sec:grover} and Figure~\ref{fig:grover}}
Additional soundness checks have to be added to the presented algorithm to validate solutions and reject invalid configurations.
These should not come at any significant runtime costs as checking if an input satisfies a Boolean formula is computationally inexpensive.
Once these invalid configurations are rejected, we can expect a uniform random distribution over multiple measurements.
With current quantum hardware and simulators, these results can only be achieved for the smallest problem instances and according circuits as the hardware requirements of the simulators (especially system memory) grow exponentially.
With this in mind, currently, we are not able to empirically confirm the approach's uniformity.

\subsubsection{Investigation of Scalability}

The width of a quantum circuit generated by our approach scales linearly in the number of features and clauses of a given CNF.
The circuit's depth (quantum runtime) is bounded by the same magnitude as Grover's search algorithm, namely $O(\sqrt{\frac{2^n}{M}})$, where $n$ is the number of features and $M$ is the size of the configuration space (number of valid configurations).
While this is clearly better than listing all valid and invalid configurations ($2^n$), it still exceeds the capabilities of current quantum hardware.
The asymptotic scaling of classical uniform random sampling algorithms is not discussed in the literature, which makes comparison difficult.

We initially assumed that our practical domain turns out to be advantageous to Grover's search as we assumed the configuration space to take a rather large (approx. $~5\%$) part of all feature combinations.
This would result in a very small number of Grover iterations necessary. 
The results show that our intuition in that regard was wrong as the potential configuration space grows much faster than the valid configuration space for all models we investigated, as we can see in Tab.~\ref{tab:fmanalysis}.
Furthermore, if the percentage of valid configurations is high, randomly guessing configurations yields valid configurations with high probability.
This could be the basis of a probabilistic sampling algorithm that counts the occurrences of randomly obtained configurations and omits already encountered (to preserve uniformity) and invalid configurations without it impacting the expected runtime too much.
However, this still requires a source of true randomness.  

Lastly, regarding scale, we did not account for the classical cost of circuit construction.
In our experiments, we were not able to construct $k=1$ circuits with more than a few hundred qubits width in Qiskit due to the used Python runtime. 
Directly generating openQASM circuits as well as further optimizations of our code base could mitigate this. We plan to investigate this as future work.

\section{Conclusion}
\label{sec:conclusion}
In this paper, we set out to answer the question "Can quantum computing improve uniform random sampling of large configuration spaces?"
To answer this question, we provided a quantum-enabled uniform random sampling method based on Grover's search algorithm that, in theory, allows us to uniformly and randomly sample configurations.
The input to our method is a feature model transformed into a CNF Boolean representation.
In the evaluation, we confirmed challenges for current NISQ-era quantum hardware to solve realistic problem instances. We discussed the feasibility, uniformity, and scalability for our approach for current quantum hardware and simulators.
To summarize, quantum computing seems a promising prospect to improve uniform random sampling of large configuration spaces, especially with true randomness. However, advances in quantum hardware are necessary for our method to compete with and potentially outperform classical uniform random sampling approaches.

\balance
\bibliographystyle{IEEEtran}
\bibliography{IEEEabrv,bibliography}

\end{document}